\documentclass[%
 reprint,
 amsmath,amssymb,
 aps,
]{revtex4-1}

\usepackage{xcolor}
\usepackage[normalem]{ulem}

\usepackage{graphicx}
\usepackage{dcolumn}
\usepackage{bm}
\usepackage{texshade}
\usepackage{amsmath}
\usepackage{amssymb}
\usepackage{mathtools}
\usepackage{float}
\usepackage{soul}

\begin{document}


\title{Dimensional confinement and superdiffusive rotational motion of uniaxial colloids in the presence of cylindrical obstacles }

\author{Vikki Anand Varma}
\affiliation{Out of Equilibrium Group, Department of Physics, Indian Institute of Technology Delhi, New Delhi 110016, INDIA}
\author{Sujin B. Babu}
\email{sujin@physics.iitd.ac.in}
\affiliation{Out of Equilibrium Group, Department of Physics, Indian Institute of Technology Delhi, New Delhi 110016, INDIA}

\date{\today}

\begin{abstract} 
In biological system like cell the macromolecules which are anisotropic particles diffuse in a crowded medium. In the present work we have studied the diffusion of spheroidal particles diffusing between cylindrical obstacles by varying the density of the obstacles as well as the spheroidal particles. Analytical calculation of the free energy showed that the orientational vector of a single oblate particle will be aligned perpendicular and a prolate particle will be aligned parallel to the symmetry axis of the cylindrical obstacles in equilibrium. The nematic transition of the system with and without obstacle remained the same, but in the case of obstacles the nematic vector of the spheroid system always remained parallel to the cylindrical axis. The component of the translational diffusion coefficient of the spheroidal particle perpendicular to the axis of the cylinder is calculated for isotropic system which agrees with analytical calculation. When the cylinders overlap such that the spheroidal particles can only diffuse along the direction parallel to the axis of the cylinder we could observe dimensional confinement. This was observed by the discontinuous fall of the diffusion coefficient, when plotted against the chemical potential both for single particle as well as for finite volume fraction. The rotational diffusion coefficient quickly reached the bulk value as the distance between the obstacle increased in the isotropic phase. In the nematic phase the rotational motion of the spheroid should be arrested. We observed that even though the entire system remained in the nematic phase the oblate particle close to the cylinder underwent flipping motion. The consequence is that when the rotational mean squared displacement was calculated it showed a super-diffusive behavior even though the orientatinal self correlation function never relaxed to zero showing this to be a very local effect.     
 
\end{abstract}

\pacs{Valid PACS appear here}
\maketitle

\section{Introduction}
The role of shape anisotropy on the diffusivity of the colloids has been studied extensively. Most of these studies are done for the bulk system, assuming particles are suspended in a medium. However, in  natural system \cite{C6CP00307A, banks2005anomalous}, the diffusivity and other dynamical behaviour changes due to many factors such as obstacles, finite size of the suspended particles, and confinement \cite{zarif2020mapping,doi:10.1021/acs.jpcb.3c00670, C3CP51585K}. It is also known that the diffusivity decreases due to the crowding effect caused by the slowly moving heavy particles\cite{sentjabrskaja2016anomalous} and also polydispersity leads to anomalous diffusion behaviour\cite{PhysRevLett.109.155901}. In some cases ions have been shown diffusing opposite to the applied potential due to the bonding with the heavy molecules \cite{lifson1962self, B711814G}. In the presence of the activity, the diffusivity of the isotropic particles has been shown to enhanced \cite{PhysRevLett.120.200606, 10.1063/1.4966188, D2SM00694D}. Local diffusivity has been found to vary with the change in the width of the channel or the radius of the obstacle \cite{doi:10.1021/j100189a004, MiguelRubi_2019}. Also, effective diffusivity has been found to reduce with the increase in the obstacle density \cite{10.1063/1.3626215, PhysRevLett.120.198103}. Diffusivity has been shown to be affected by the presence of the effective force field, e.g., gravity \cite{PhysRevLett.115.020601}. In the passive particle system, diffusivity has been shown to decrease with the increase in obstacle concentration \cite{doi:10.1021/acsomega.3c05492}, and the nature of glass transition at high density has also been studied \cite{PhysRevE.82.041505}. In the $2D$ system, the presence of obstacles leads to sub-diffusive behavior \cite{PhysRevE.89.022708,doi:10.1021/acsomega.3c05945} for the Brownian particle. The $2D$ confinement also leads to the anomalous behaviour, in the presence of the obstacles \cite{PhysRevLett.92.250601}. The confinement and the presence of obstacles leads to phase rich behaviours by changing the structural properties. In the presence of solute-solvent attractive interaction, colloids have been observed altering the alignment of the nematic fluid \cite{PhysRevE.102.062705}. The liquid crystals have been found aligning with the surface of the large spheres sandwiched between the two parallel surfaces \cite{PhysRevLett.85.4719}, forming a Saturn ring kind of structure. Study of such systems also opens a pathway for the application in various engineering solution e.g, in electronics.

These observed effects depend upon the type of confinement \cite{PhysRevE.107.044604}, absorption properties of the confining walls \cite{10.1063/5.0164257}, the surface roughness of the wall\cite{YEH2021110282}, drift forces \cite{10.1063/1.3651478} etc . Apart from these physical and chemical properties of fluid and confining geometry, the dynamics of colloidal particle also depend upon its shape \cite{doi:10.1021/acs.langmuir.0c01884}, especially in the active particle system \cite{10.1063/1.5081507}. For a 2D system, having confinement along one axis, the system has been shown to follow the behavior of a 1D system\cite{zarif2020mapping}. In these system, anisotropic particles can behave differently than isotropic particle system. In the presence of the anisotropic particles, the role of obstacles has been paid less attention, specifically the study of the dynamics and structure while considering interaction of the phase-rich anisotropic particles with obstacles. 

These systems have been studied extensively by using both analytical and experimental approaches. For example, the mean first passage time \cite{guerin2016mean}, is shown to be an important parameter governing the kinetics of the system, especially for diffusion-limited cluster aggregation (DLCA) processes. Another example, the diffusivity of the polymer has been studied using blob theory \cite{PhysRevLett.110.168105}, an approximation done in terms of average particle size. In the presence of periodic obstacles analytical relations have been derived  for the diffusion coefficient considering them as point particles\cite{SIMPSON20173346, farah2020diffusion}. The effective diffusivity in the dimensional confinement for the $2$ dimensional system has been shown for the point particles \cite{10.1063/5.0009095}, where obstacle configuration leads particles to move within a channel. Few lattice-based models are also proposed for the study of the diffusion of the particle with obstacles \cite{ellery2015calculating}. In most of the studies performed previously on similar kind of system, the radius of the obstacles considered is much bigger than the size of the Brownian particles \cite{10.1063/1.4720385}. In all these studies the shape anisotropy of the particles were not considered. In a case, where analytical approaches becomes very difficult to follow, simulation can be a useful tool. In simulation, different coarse-grained models provide an excellent advantage for studying comparatively complex systems. The anomalous diffusivity in the presence of crowded media has been studied in the periodic obstacles where the motion becomes diffusive in the long time limit \cite{C4CP03599B,sentjabrskaja2016anomalous}. 

In the present work we have considered a system of anisotropic particles in the presence of the periodic arrangement of the cylindrical obstacles. Depending upon the density of the obstacles system is effectively confined. For example at the high density of cylindrical obstacles \cite{10.1063/1.3651478} particles are constrained to diffuse in one dimension. In this way, at different obstacle configurations, we studied the structure, kinetics and dynamics emerging in such phase rich system of hardcore particles.

For highly elongated particles, the system is well known to encounter nematic alignment. In our case, the director axis has been found to align with the obstacles. In the present study we also observed that the rotational dynamics of the oblate particles are not freezing despite being in the nematic phase.  Nematic-isotropic transition lines have been found to shift in the two-component system \cite{PhysRevE.106.014602}. In the presence of confinement the orientation of the nematic director could be controlled as well, for example hard rods have been found to align along the plane of the substrate, shown using density functional theory \cite{PhysRevA.38.3721}. Alignment of the particles depending upon the boundary of the medium has been found to govern the crack patterns\cite{PhysRevE.86.016114, PhysRevE.59.1408}, growth of bacteria\cite{langeslay2024stress} etc. The presence of obstacles has also been found to affect the self-assembly of the Brownian particles \cite{C5SM01952D,Shireen_jcp_2017,shireen_sf_2018}. The presence of different types of particles or obstacles also alters the phase boundary, as shown in the binary mixture of hard-spheroid-sphere and sphere-rod systems \cite{adams1998entropically,10.1063/1.1507112}. However in the presence of the cylindrical obstacles, despite governing the orientation of the nematic director the system shows no alteration in its nematic-isotropic phase boundary. These kinds of behaviour can help us to design colloidal systems with a robust control \cite{design} over structural properties for a variety of application.

\section{Simulation method}
\label{sm}
We start the simulation by randomly distributing $N$ number of particles in a cubic box with edge length $L$, where all the lengths have been considered in the unit of $d=1$, the diameter of the spherical Brownian particle. Thus volume fraction of the particles is defined over the available volume ${V-\pi r_\circ^2 L}$ such that  $\phi=\frac{N}{V- \pi r_\circ^2 L}$, where $V=L^3$. The volume of the anisotropic particle with the major axis of length $a$ and minor axis length $b$ having aspect ratio $p=a/b$ is always kept equal to the volume of a sphere with diameter $d=1$. Thus the volume fraction of the spheroidal particles is given by $\pi/6 a b^2$. The shape anisotropic spheroidal particle orientation vector $\hat{n}$ has been considered along its symmetry axis. The number of particles considered are in the range of $500-2500$ and $p$ is simulated in the range $0.25-4$. The cylindrical obstacles of radius $r_\circ$ and length $L$ are placed parallel to each other along the $z$ direction on a square lattice. Periodic boundary condition is applied in the cubic box. The density of the obstacles is defined in terms of the area fraction $\phi_\circ$ calculated in the plane perpendicular to the axis of cylinders. Where $\phi_\circ = \rho_\circ \pi r_\circ^2$, such that $\rho_\circ$ is the number fraction of the obstacles defined as $\rho_\circ=\frac{N_\circ}{L^2}$. 

To perform the dynamics of the particles, we used the Brownian-cluster-dynamics (BCD) simulation technique \cite{babu_jcp_2006}. Using this method structure, kinetics and dynamics for a variety of different types of Brownian particle like sphere \cite{zakiya_acso_2023,malhotra_jcp_2019,yadav_jml_2024}, spheroids \cite{varma_ats_2023,varma_pccp_2024} are studied. In this method, we randomly select particles $2N$ times and perform either a rotational or translational movement with equal probability. It ensures that each particle is translated or rotated at least once, in each simulation step. For the translational movement, we displace the particle in a random direction with a fixed step length $S_T$. Similarly, for the rotational motion, the tip of the orientation vector $\hat{n}$ performs a $2d$ random walk over the surface of a sphere, with a fixed rotational step length $S_R$, such that $S_T$ and $S_R$ satisfy the following relation,
\begin{equation}
    S_R^2 = 2 S_T^2.
\label{eq.4_1}
\end{equation}
The relation between the simulation time $t_{Sim}$ and physical time $t_{Phy}$ is given as,
$\frac{t_{Phy}}{t_0} = t_{Sim}\frac{{S_T }^2}{d^2}$ \cite{PhysRevE.106.014602}. Where $t_\circ$ is the time required for a single sphere to diffuse through a distance of its own diameter.
After performing each movement step, we check the overlap condition of the hardcore particle and if the particles are overlapping then we reject that particular movement \cite{PhysRevE.106.014602}.

\subsection{Dynamics of the shape anisotropic particles}

To implement the dynamics for the shape anisotropic particles, we resolve $S_T$ along $\hat{n}$ $S_T^\parallel$ and perpendicular to $\hat{n}$ \cite{PhysRevE.106.014602}. $S_T^\parallel$ as given below, 

\begin{subequations}
\begin{align}
\frac{ S_T^{\perp , \parallel} }{S_T} = \sqrt{ \frac{d}{2b G_T^{\perp or \parallel}} }  \\ 
\frac{ S_R^e }{S_T} =  \sqrt{\frac{2}{G_\theta} } 
\end{align}
\label{eq.4_2}
\end{subequations}
Where $G_T^{\perp or \parallel}$ is the Perrin's friction factor for the diffusion of the particle perpendicular and parallel to $\hat{n}$, and $G_\theta$  is the friction factor for the rotational motion, which is calculated analytically for the stick boundary condition, for the prolate ($p>1$) case,

\begin{subequations}
\begin{align}
G_T^{\parallel}= \frac{4}{3} \left[ \frac{p}{ \left( 1-p^2 \right) } + \frac{ 2p^2-1 }{ \left(p^2-1 \right)^{3/2} } \ln \left(p+\sqrt{p^2-1}\right) \right]^{-1}  \\
G_T^\perp= \frac{8}{3} \left[ \frac{p}{ \left( p^2-1 \right) } + \frac{ 2p^2-3 }{ \left(p^2-1 \right)^{3/2} } \ln \left(p+\sqrt{p^2-1}\right) \right]^{-1} \\
G_\theta= \frac{2}{3} \frac{ \left(p^4-1 \right) }{p} \left[ \frac{ \left(2p^2-1 \right) }{ \sqrt{p^2-1}} \ln \left(p+\sqrt{p^2-1} \right) -p \right]^{-1}.
\end{align}
\label{eq.4_3}
\end{subequations}

and for the oblate ($p<1$) case, we have,
\begin{subequations}
\begin{align}
G_T^{\parallel}= \frac{4}{3} \left[ \frac{p}{ \left( 1-p^2 \right) } + \frac{ 1-2p^2 }{ \left(1-p^2 \right)^{3/2} } \arccos \left(p \right) \right]^{-1}  \\
G_T^\perp= \frac{8}{3} \left[ \frac{p}{ \left( p^2-1 \right) } + \frac{ 3-2p^2 }{ \left(1-p^2 \right)^{3/2} } \arccos \left(p \right) \right]^{-1} \\
G_\theta= \frac{2}{3} \frac{ \left(p^4-1 \right) }{p} \left[ \frac{ \left(2p^2-1 \right) }{ \sqrt{1-p^2}} \arccos \left(p \right) -p \right]^{-1}
\end{align}
\label{eq.4_4}
\end{subequations}
The parameter $G$ is $1$, for $p=1$ that means the particle is spherical.



\subsection{Surface energy calculation close to the obstacle-particle interface}
The grand thermodynamical potential $\Omega$ can be expressed as a functional of density $\rho$ \cite{poniewierski1988nematic, PhysRevA.38.3721},
\begin{equation}
    \Omega\{\rho(\mathbf{r}, \omega)\}=F_{id}+F_{f}+ F_{V} + F_\mu
\label{eq.4_5}
\end{equation}
where $F_{id}$  is the ideal gas contribution give as, 
\begin{equation}
F_{id} = \int \left[  \ln{\Lambda^3 \rho (\mathbf{r},\omega)} -1 \right] \rho(\mathbf{r}, \omega)  \,d\omega \,d\mathbf{r}
\label{eq.4_6}
\end{equation}

Inter-particle interaction is counted in the term $F_{f}$  given as,

\begin{equation}
    F_{f}= \int  f(\mathbf{r}_1, \mathbf{r}_2,  \omega_1, \omega_2) \rho(\mathbf{r_1}, \omega_1) \rho(\mathbf{r}_2, \omega_2) \,d\omega_1 \,d\omega_2 \,d\mathbf{r}_1 \, d\mathbf{r}_2
\label{eq.4_7}
\end{equation}

where $ f(\mathbf{r}_1, \mathbf{r}_2,  \omega_1, \omega_2)$ is the mayer function. This function has a value $1$ for overlap and $0$ for non-overlapping conditions.
$F_\mu$ is the potential due to the hard core contribution, given as 

\begin{equation}
    F_\mu=-\beta \mu  \int \rho(\mathbf{r}, \omega) \, d\omega \,d\mathbf{r} 
\label{eq.4_8}
\end{equation}
 where $\mu$ is the chemical potential of the system at a particular volume fraction in the absence of the obstacles.
$F_V$ is the free energy contribution in the presence of a cylinder given as,

\begin{equation}
    F_V=\int V_o(\mathbf{r}, \omega) \rho(\mathbf{r}, \omega) \, d \omega \, d \mathbf{r}
\label{eq.4_9}
\end{equation}

We express $\mathbf{r}$ in cylindrical co-ordinate as, $\mathbf{r}=\varrho \hat{\varrho} + z \hat{z}$. In Cartesian co-ordinate, it can be written as $\mathbf{r}=\varrho \hspace{0.2 cm} cos(\theta)\hat{i}+ \varrho \hspace{0.2 cm} sin(\theta) \hat{j}+ z \hat{k}$.

We express the potential in the presence of cylindrical obstacles as, 
\begin{equation}
V_o=
\begin{cases}
+ \infty \hspace{0.5cm} \varrho < \varrho_m(\theta, \omega, r_\circ) \\
0 \hspace{0.5cm}  \varrho > \varrho_m(\theta, \omega, r_\circ)
\end{cases}
\label{eq.4_10}
\end{equation}

where $r_\circ$ is the radius of the cylindrical obstacles and $\varrho_m$ is the minimum distance between the center of the obstacle and the center of mass of the Brownian particle, which depends upon the relative orientation of the particle and the cylinder body. For a particular system, since we have  $r_\circ$ as constant, $V_\circ$ can be expressed as,

 \begin{equation}
    V_o=
\begin{cases}
+ \infty \hspace{0.5cm} \varrho < \varrho_m(\theta, \omega) \\
0 \hspace{0.5cm}  \varrho > \varrho_m(\theta, \omega)
\end{cases} 
\label{eq.4_11}
\end{equation}

We minimize $\Omega\{\rho(\mathbf{r}, \omega)\}$ with respect to $\rho(\mathbf{r}, \omega)$ which leads to,
\begin{equation}
    \rho(\mathbf{r}, \omega)= \rho_{b}(\mathbf{r}, \omega) \exp{[-V_o(\varrho, \theta, \omega)]}
\label{eq.4_12}
\end{equation}
Where $\rho_b(\mathbf{r}, \omega)$ is the bulk properties of the Brownian particles that can be retrieved when $V_o$ is zero.
\begin{align}
\rho_{b}(\mathbf{r}, \omega) =  \frac{1}{\Lambda} e^{\beta\mu} \hspace{4 cm}\nonumber  \\ 
\times \exp{[\int f(\mathbf{r},\mathbf{r}',\omega', \omega) \rho(\mathbf{r}, \omega) \rho(\mathbf{r}', \omega) \, d \mathbf{r}' \, d\mathbf{\omega}']} 
\label{eq.4_13}
\end{align}
The term that represents the bulk property can also be written as, $\rho_b(\mathbf{r}, \omega)= \rho_b f(\omega)$ \cite{poniewierski1988nematic} assuming uniform distribution of the particles. Where $f(\omega)= f(\mathbf{n_1}.\hat{n})$ is the angular distribution of particle calculated from the director of the nematic-alignment $\mathbf{n}_1$. $f(\omega)$ can be normalized with $4\pi$ in the case of isotropic phase.
To calculate the system's energy close to the obstacles, we consider the symmetry of the cylindrical obstacle along $\hat{z}$. Now, $\rho$ can be expressed as $\rho{(\varrho, \theta, \omega)}$ given by,

\begin{equation}
\rho(\varrho, \theta, \omega) =\rho_b f(\omega) \hspace{0.2 cm} \exp{[-V_o(\varrho, \theta, \omega)]}
\label{eq.4_14}
\end{equation}
To calculate the surface energy per unit length along the symmetry axis of the obstacle, we take both the densities, bulk, and close to the surface and subtract the energy calculated by using eq. \ref{eq.4_5}, eq. \ref{eq.4_10} and eq.\ref{eq.4_11}. The surface energy per unit length, $\psi_l$ is given as,

\begin{align}
\frac{\psi_l}{k_BT}=-\frac{S}{k_B}-\frac{\Delta\mu}{k_BT} \hspace{0.1 cm} \xi
\label{eq.4_15}
\end{align}
 \begin{equation}
     \Delta\mu=\mu-k_BT \ln{(\Lambda^3 \rho_b/4\pi)}
    \label{eq.4_16}
 \end{equation}

\begin{align}
    \xi =  \int_{\theta=0}^{2\pi} \int_{\omega=0}^{\pi} \int_{\varrho=r_\circ}^{\infty}[\rho(\varrho, \theta, \omega)-\rho_b f(\omega)] \, d \varrho \, \varrho \, d \theta \, d\omega  \nonumber \\ 
    =- \frac{1}{2} \rho_b \int_{\theta=0}^{2\pi} \langle  (\varrho_m^2(\omega, \theta)-r_\circ^2) \rangle_{f(\omega)} \,d\theta \hspace{0.5cm}
\label{eq.4_17}
\end{align}

The other term contributing to the surface energy per unit length contains the rotational term coming from the anisotropic configuration of the particle and is given as,

\begin{equation}
    S_{rot}= \frac{1}{2} \rho_b \int_{\theta=0}^{2\pi} \langle (\varrho_m^2(\omega, \theta)-r_\circ^2)  \ln{[4\pi f(\omega)]} \rangle_{f(\omega)}\, d\theta
\label{eq.4_18}
\end{equation}

Where $S$ is given as,
\begin{equation}
S=S_{ro}+S_{tr}
\label{eq.4_19}
\end{equation}
, such that $S_{tr}= S_{tr}^{id}+S_{tr}^{m}$
Where the ideal and  inter-molecular term can be defined as,

\begin{align}
S_{id}=-\frac{1}{2} \rho_{b} \int_{\theta=0}^{2\pi} \langle (\varrho_m^2(\omega, \theta)-r_\circ^2)  \rangle_{f(\omega)} \, d \theta
\label{eq.4_20}
\end{align}

\begin{align}
S_{tr}^{m}=-\frac{1}{2}[\int f_2(\boldsymbol{\varrho}, \omega, \mathbf{r}', \omega' ) \rho(\boldsymbol{\varrho}, \omega) \rho(\mathbf{r}', \omega') \, d\mathbf{r}' \, d\omega' \, d\boldsymbol{\varrho} \, d\omega \nonumber  \\
-\rho_b^2 \int f_2(\boldsymbol{\varrho}, \omega, \mathbf{r}', \omega' )  f(\omega) f(\omega') \, d\mathbf{r}' \, d\omega' \, d\boldsymbol{\varrho} \, d\omega]
\label{eq.4_21}
\end{align}

\begin{figure}
\includegraphics[height=5.25cm,width=8.5cm]{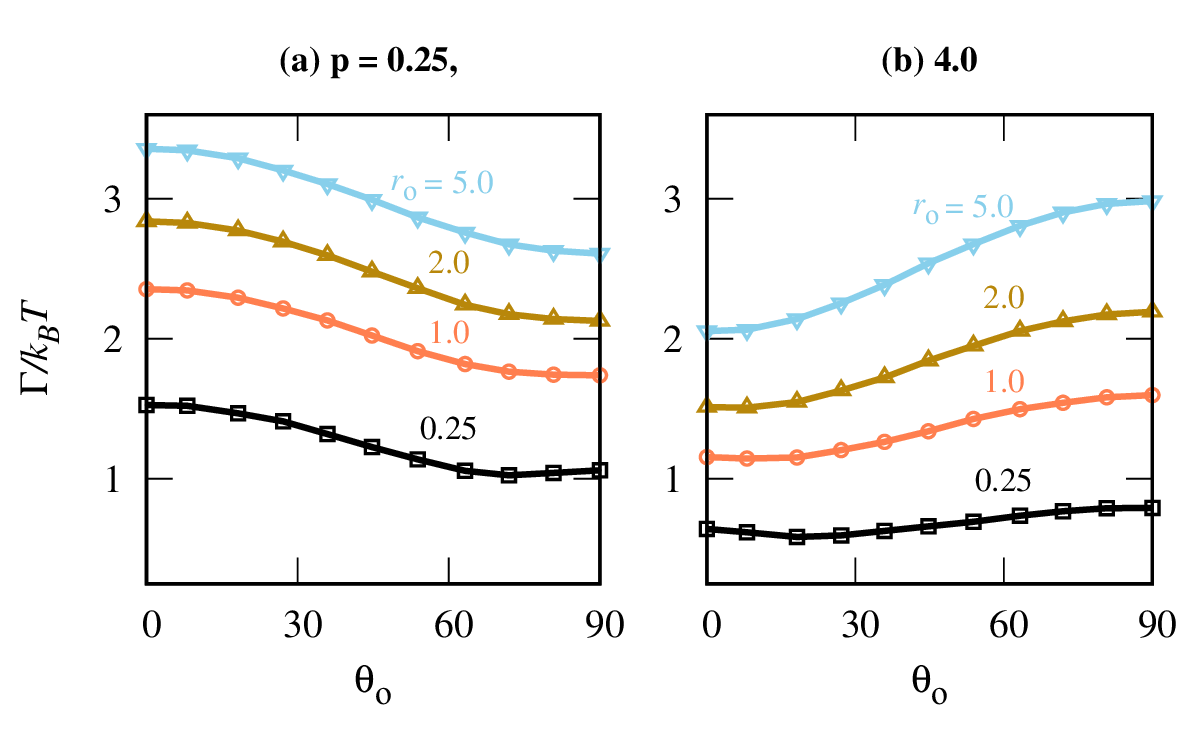}
  \caption{Surface energy is calculated at different $\theta^\circ$, for both the oblate ($p=0.25$) and prolate ($p=4.0$) system. Shown for the different radii of the cylindrical obstacles, as depicted in the plots. Oblate 
 and prolate systems are energetically favored at $\theta^\circ=90$ and $0$ respectively.}
\label{fig.4_1}
\end{figure}

\begin{figure}
\includegraphics[height=2.9cm,width=8.5cm]{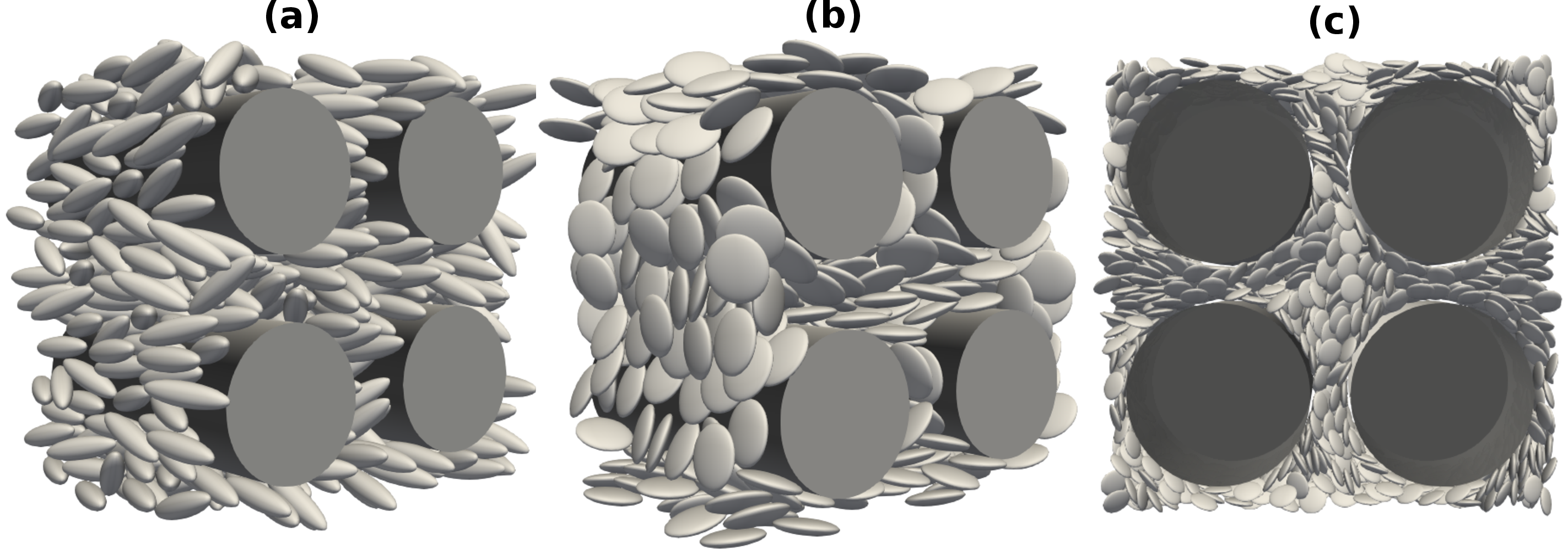}
  \caption{Snapshot of the system in nematic-phase, is shown at $\phi=0.45$ for (a) $p=4.0$, $r_\circ=2.0$ and $\phi_\circ=0.4$, (b) $p=0.25$, $r_\circ=2.0$ and $\phi_\circ=0.4$,  (c) $p=0.25$, $r_\circ=5.0$ and $\phi_\circ=0.55$, where all the three systems are in nematic phase.}
\label{fig.4_2}
\end{figure}

The average surface energy per unit area at the obstacle bulk interface can be expressed as, 

\begin{align}
\frac{\Gamma}{k_BT}=\frac{3a}{2\pi [(3a/2+r_\circ)^2- r_\circ^2]}\frac{\psi_l}{k_BT}
\label{eq.4_22}
\end{align}

Here the greater value of $\mu$ contributes to the alignment of the particles along the cylinder's axis. In other words, for a higher $\phi$ region, where $\mu$ is greater the parallel alignment is ensured. We took a value of $\mu$ close to the isotropic-nematic interface and calculated eq. \ref{eq.4_16}, eq.\ref{eq.4_17}, eq. \ref{eq.4_18}, eq. \ref{eq.4_20} and eq. \ref{eq.4_21}.  Where eq. \ref{eq.4_21} was calculated by using Monte-Carlo and the curve-fitting method, as the function is smooth all over the region of integral. Fig. \ref{fig.4_1} shows the variation in the $\Gamma/k_BT$ with the angle between the orientation of the director and the axis of the cylinder, $\theta_\circ$. For the oblate case ($p=0.25$) the angle perpendicular to the cylinder's axis is favored, over the parallel arrangement. In the case of prolate ($p=4.0$), we have parallel alignment favored over the perpendicular alignment. The energy difference between parallel and perpendicular alignment reduces as the radius of the obstacle $r_\circ$ decreases. For the thin cylinders, the energy minima occur slightly away from the $\theta_\circ=90^\circ$ in the oblate and $0^\circ$ in the prolate case. As we increase $r_\circ$, the difference between $\Gamma /k_BT$  of minimum and maximum energy configuration increases. In other words, the control of the nematic axis is enhanced with the increase in the radius of the cylinder \cite{medlin2009interfaces} at a constant number density of the obstacles.
\section{Results}
\label{result}
\subsection{Kinetics of the evolution of nematic-alignment in the presence of the cylindrical obstacles}

\begin{figure}
\includegraphics[height=4.25cm,width=8.5cm]{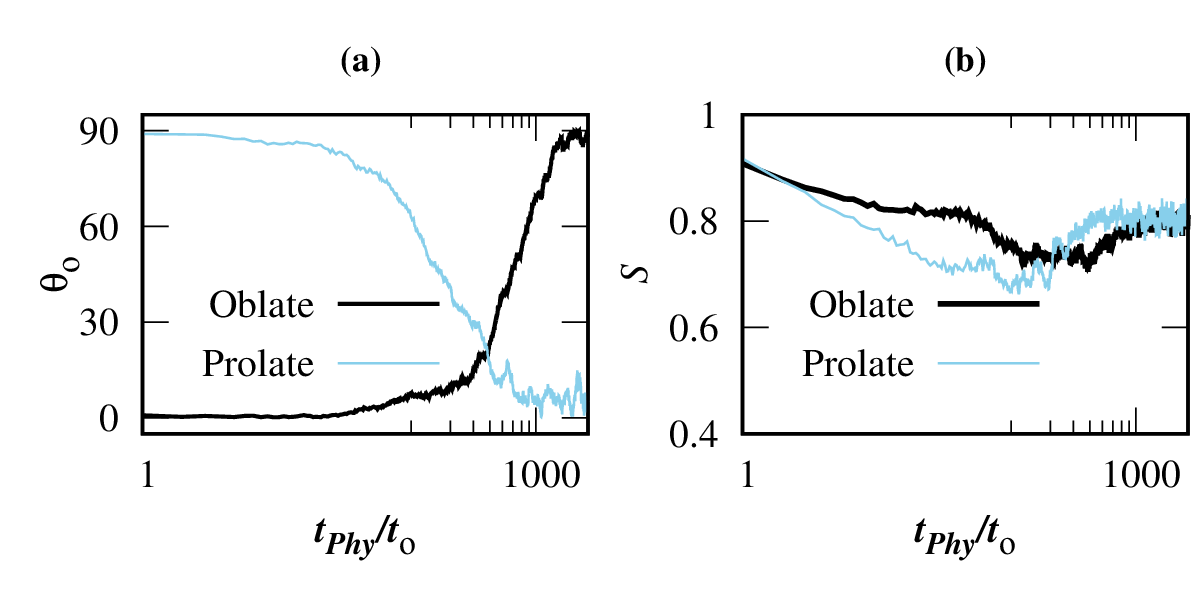}
  \caption{(a) Evolution of the director of oblate and prolate system is shown for $r_\circ=2.0$, and a low $\phi_\circ=0.063$, at $\phi=0.5$. The system starts initially from an unfavoured condition for prolate ($p=4.0$) at $\theta_\circ = 90^\circ$, oblate ($p=0.25$) at $\theta_\circ=0^\circ$, where $\theta_\circ$ is the angle between the nematic director and the axis of the cylindrical obstacles. (b) The evolution of the order parameter is shown for both oblate and prolate systems. Both the systems show stability after a long time limit($t_{phy}>1000$).}
\label{fig.4_3}
\end{figure}

\begin{figure}
\includegraphics[height=10.9cm,width=8.7cm]{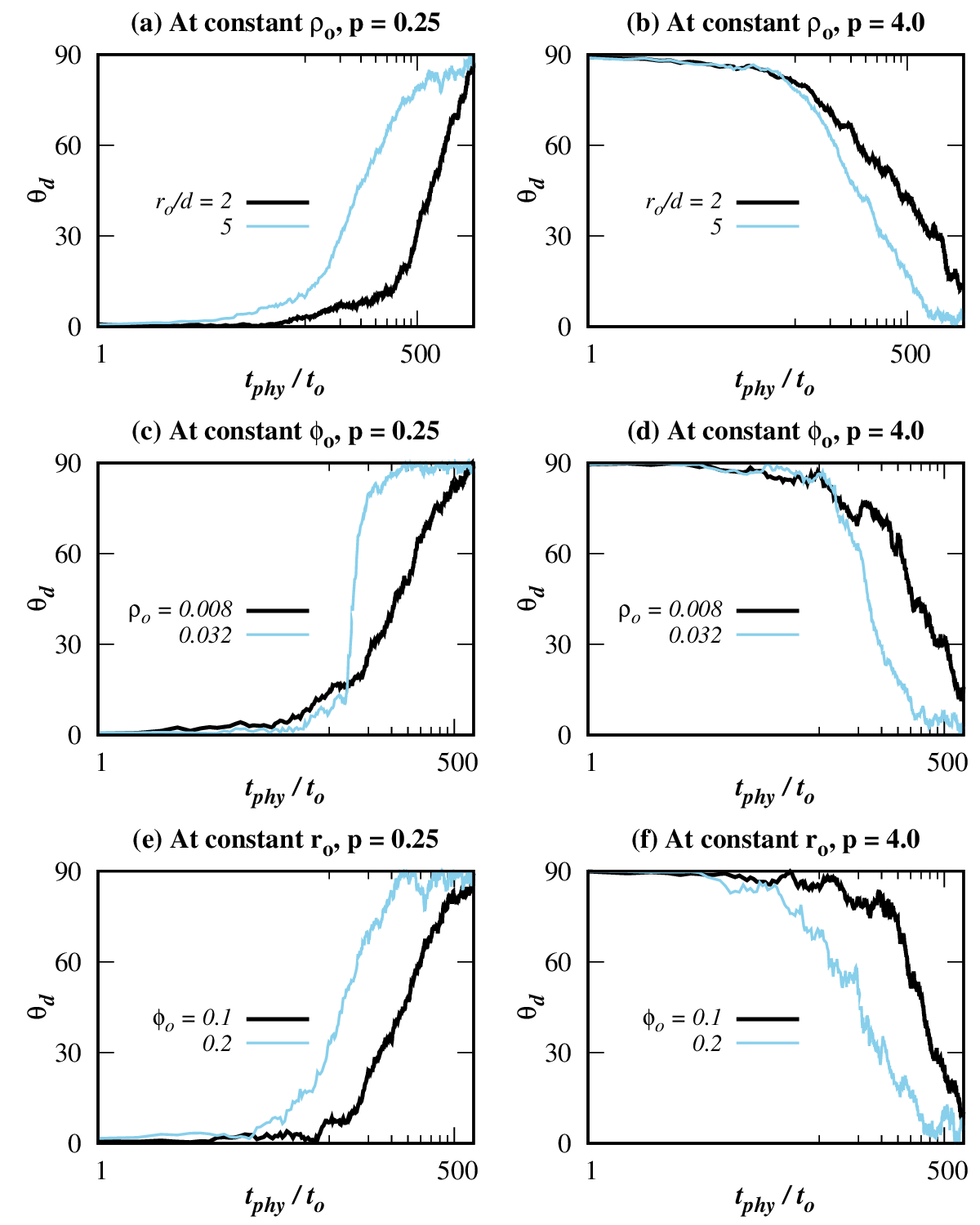}
  \caption{Evolution of the directors shown at the different configuration for both the oblate and prolate system. (a)Prolate and (b) oblate are shown at $\phi=0.5$ for the different $r_o/d$, at a constant $\rho_{o}=0.005$. (c) and (d) shows the evolution of the oblate and prolate system for different $\rho_o$ and at a constant $\phi_o=0.1$. (d) and (f) shows the system's evolution for different $\phi_o$, at a constant $r_o=2.0$.}
\label{fig.4_4}
\end{figure}

\begin{figure}
\includegraphics[height=8.7cm,width=8.7cm]{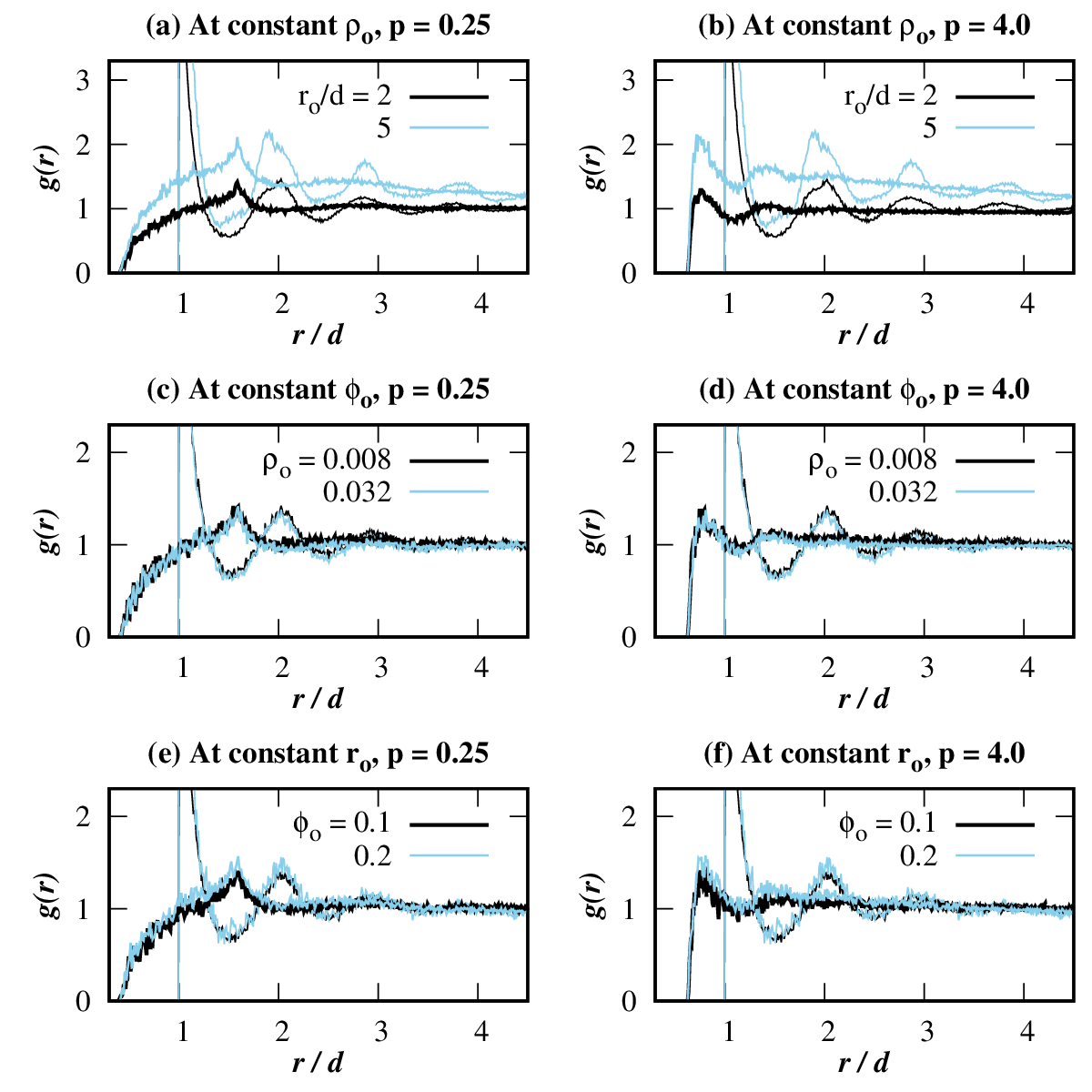}
  \caption{$g(r)$ is shown for the system in the nematic phase for both prolate and oblate particles. (a) and (b) at constant $\rho_\circ=0.005$, (c) and (d) at constant $\phi_\circ=0.1$, (e) and (f) at constant $r_\circ=2.0$. Where thin lines show the system of spheres equilibrated at the same configuration.}
\label{fig.4_5}
\end{figure}

\begin{figure}
\includegraphics[height=8.7cm,width=8.7cm]{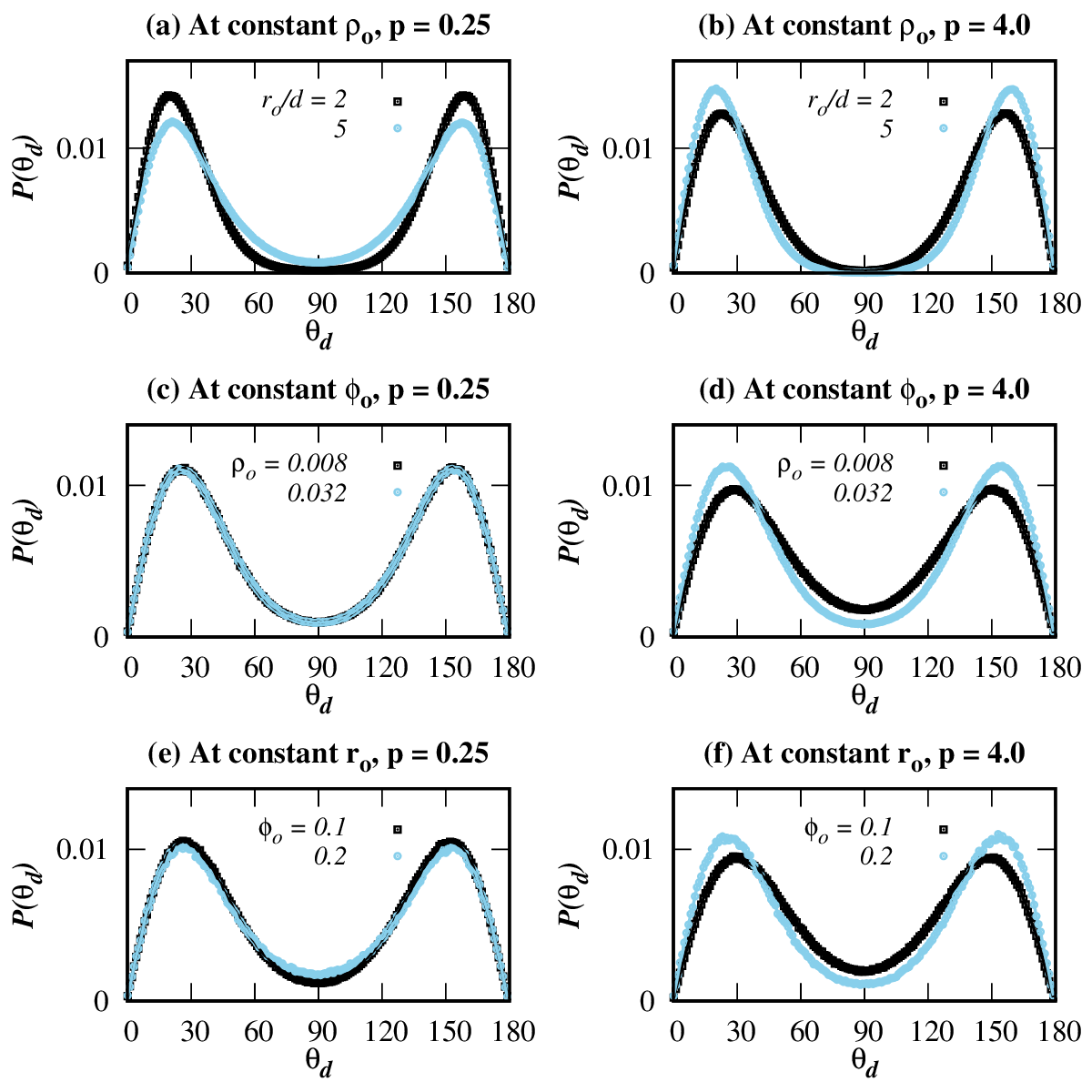}
 \caption{$g(\theta)$ is shown for the system in the nematic phase for both prolate and oblate particles. (a) and (b) at constant $\rho_\circ=0.005$, (c) and (d) at constant $\phi_\circ=0.1$, (e) and (f) at constant $r_\circ=2.0$. }
\label{fig.4_6}
\end{figure}

Fig. \ref{fig.4_1} shows the surface energy per unit volume close to the cylinder-bulk interface as given by eq. \ref{eq.4_22}, where we considered only the hardcore repulsion interaction between the cylinder-ellipsoid and ellipsoid-ellipsoid. Also, we have considered only a single cylindrical obstacle. In fig. \ref{fig.4_1}(a), we observe that the surface energy is minimum when the symmetry axis of the oblate particle ($p=0.25$) is perpendicular to the cylindrical axis. For the prolate particle ($p=4.0$), we observe that minimum surface energy configuration happens when the symmetry axis of the particle is parallel to the cylindrical axis. Thus, it seems that the presence of an obstacle will assist in the nematic transition for anisotropic particles. However, we would like to mention that the nematic-isotropic phase transition was same as the system without obstacles, even for the high density limit at various obstacles configuration as well as for different aspect ratio particles as well.   

We initiate the system by putting The spheroids in an energetically less favorable state as obtained from obstacle-bulk surface energy calculation (see fig.\ref{fig.4_1}). For the oblate particle, nematic director is less favoured along the symmetry axis of the cylinders and for the prolate case it is less favoured along the perpendicular plain of the symmetry axis of the obstacles. Thus the ellipsoids in the initial configuration are already in the nematic phase. The relaxation of the angle between the symmetry axis of the ellipsoid and the cylindrical axis for $\phi=0.5$, $\rho_\circ=0.005$ and $r_\circ=2.0$ can be observed in Fig. \ref{fig.4_3}(a). It seems that the rotational kinetics of the symmetry axis of the ellipsoids relaxes slowly up to a time of $200$, and then suddenly, all the ellipsoids rearrange to the minimum surface energy configuration (see Fig \ref{fig.4_2}(a) and Fig \ref{fig.4_2}(b)) obtained from a single ellipsoidal calculation. Also note that in the absence of obstacles at $\phi=0.5$, the equilibrium configuration of the ellipsoids is a nematic phase. Fig. \ref{fig.4_3}(b) shows the evolution of the nematic phase order parameter $S$, which is the largest eigenvalue of the tensor $Q$, defined by,
\begin{equation}
Q_{\alpha,\beta} = \frac{3}{2} \frac{1}{N} \sum_i \langle(n_\alpha)_i (n_\beta)_i \rangle - \frac{1}{2} \delta_{\alpha,\beta}
\label{eq.19}
\end{equation}
where $n_\alpha$ and $n_\beta$ are the components of the particle's orientation vector$\hat(n)$ with $\alpha, \beta \in{x,y,z} $. 


As the system moves towards the equilibrium state, the $S$ parameter falls at initial times but remains well above the value $0.3$, showing that the system never passes through the isotropic phase. In other words, the system remains in the nematic phase throughout the evolution in both prolate and oblate cases. The fall in the $S$ parameter indicates the gradual change in the orientation of the particles, which is initiated from the cylinder-bulk interface and expands towards the bulk region of the ellipsoidal particles. This shows that we can make an ellipsoid particle system undergo a transition from an unfavorable nematic phase to a favorable nematic phase due to the presence of the cylindrical obstacle.


Fig. \ref{fig.4_4} shows the kinetics of the evolution of $\theta_\circ$ for both oblate and prolate cases in the presence of obstacles. In the case of constant $\rho_\circ=0.005$, the kinetics of the system is higher for the system with cylinders having a greater radius, as shown in the Fig. \ref{fig.4_4} (a) and \ref{fig.4_4} (b).  A system with  $r_\circ=5$ evolves faster than a system with $r_\circ=2$. It can be explained by considering the effect of $\phi_\circ = \rho_\circ \pi r_\circ^2$, such that $\phi_\circ$ increases, with the increase in $r_\circ$. A system with a higher surface of the obstacles, will have a higher total surface energy difference between favorable and unfavorable states, which results in faster evolution towards the favorable state. Apart from that, $\Delta\Gamma_m$ has higher value for the obstacles having greater $r_\circ$, this effect is not as dominating as the effect of obstacles-bulk interface interaction. It has been shown in the Fig. \ref{fig.4_4} (c) and \ref{fig.4_4} (d), where both the system was simulated at constant $\phi_\circ=0.1$, the system with higher $\rho_\circ=0.032$  shows faster evolution to the favorable state. 

At a constant $\phi_\circ$, reducing the radius of the cylinders will increase the surface area as the number of cylinders will increase in the system.  In other words, the sky-blue line representing a system with $r_\circ=1.0$ has a surface area of the obstacles two times greater than the system with $r_\circ=2.0$, at constant $\phi_\circ$. Therefore, despite having $\Delta\Gamma_m$ higher, the system with $r_\circ=2$ shows slower evolution than the system with $r_\circ=1$. Similar observations are shown in Fig. \ref{fig.4_4} (e)  and \ref{fig.4_4} (f) for the oblate and prolate systems, respectively.  For the same $r_\circ$, the system with higher $\phi_\circ$ has a greater surface area in contact with the ellipsoids, and therefore system with higher $\phi_\circ$ evolves faster.  

To show the structural properties of the nematic phase, we calculated the radial distribution function $g(r)$, shown in Fig. \ref{fig.4_5}. For $r_\circ = 5$ and $\rho_\circ=0.005$, we have $g(r)$ curve deviating from the usual $g(r)=1$ (solid line), due to the absence of the particles in the obstacle regions, for both oblate and prolate cases, which is shown in the Fig. \ref{fig.4_5} (a) and \ref{fig.4_5} (b) respectively. The dotted line represents the system with spherical particles simulated at the same configuration. However, the peaks of the curves are similar to the bulk system, where again the deviation represents a large inaccessible region for the diffusing particles. The black lines represent the system with $r_\circ=2$ simulated at the same $\rho_\circ=0.005$, which show characteristics similar to the bulk phase. Where we have a small inaccessible region in comparison to the $r_\circ=5$. For both systems, the local structure remains similar to the bulk phase. Fig. \ref{fig.4_5} (c) and \ref{fig.4_5} (d) represent the system of oblate and prolate particles, respectively, simulated at constant $\phi_\circ=0.1$ for different obstacle radius, where sky-blue lines represent $r_\circ=1$ and black with $r_\circ=2$. The system, again, can be observed to be similar to the bulk phase. Fig. \ref{fig.4_5} (e) and \ref{fig.4_5} (f) shows the $g(r)$ for the system of oblate and prolate particles, respectively, simulated at constant $r_\circ=2$ for different obstacle density. Where sky-blue lines represent $\phi_\circ=0.2$ and black lines $r_\circ=0.1$. 

To understand the alignment of the particles in the nematic phase, we have shown $g(\theta)$ as a function of the angle $\theta$, where $g(\theta)$ is the distribution of the angle $\theta$, formed between $\hat{n}$ of any two particles. For a perfectly aligned nematic system, it will show sharp peaks at $\theta=0^\circ$ and $ 180^\circ$. Fig. \ref{fig.4_6} (a) and Fig. \ref{fig.4_6} (b) shows the distribution $g(\theta)$ for prolate and oblate particles respectively, at a constant $\rho_\circ=0.005$ at two different obstacle radius $r_\circ=2$ and $r_\circ=5$. For the oblate case Fig. \ref{fig.4_6}(a), there is less alignment with increasing the radius of the obstacles. It has effect on the dynamics of the system, where it shows an anomalous rotational behaviour, which has been discussed in the later section of this work.  While, for the prolate, the fraction of particles aligned increases, with the radius of the obstacles. Fig \ref{fig.4_6}(c) and Fig \ref{fig.4_6}(d), shows the alignment calculated for the same $\phi_\circ$, where we observe increasing the number density while decreasing the radius of the cylinders leads to better alignment in the case of prolate particles, while for the case of oblate particle, it remains nearly the same. With increasing $\phi_\circ$, we observed a small decrease in alignment for the oblate particles see Fig. \ref{fig.4_6}(e), where as prolate shows better alignment see Fig . \ref{fig.4_6}(f). In all three cases, the behavior of oblate has been seen to be opposite to the prolate case. 

\begin{figure}
\includegraphics[height=2.9cm,width=8.6cm]{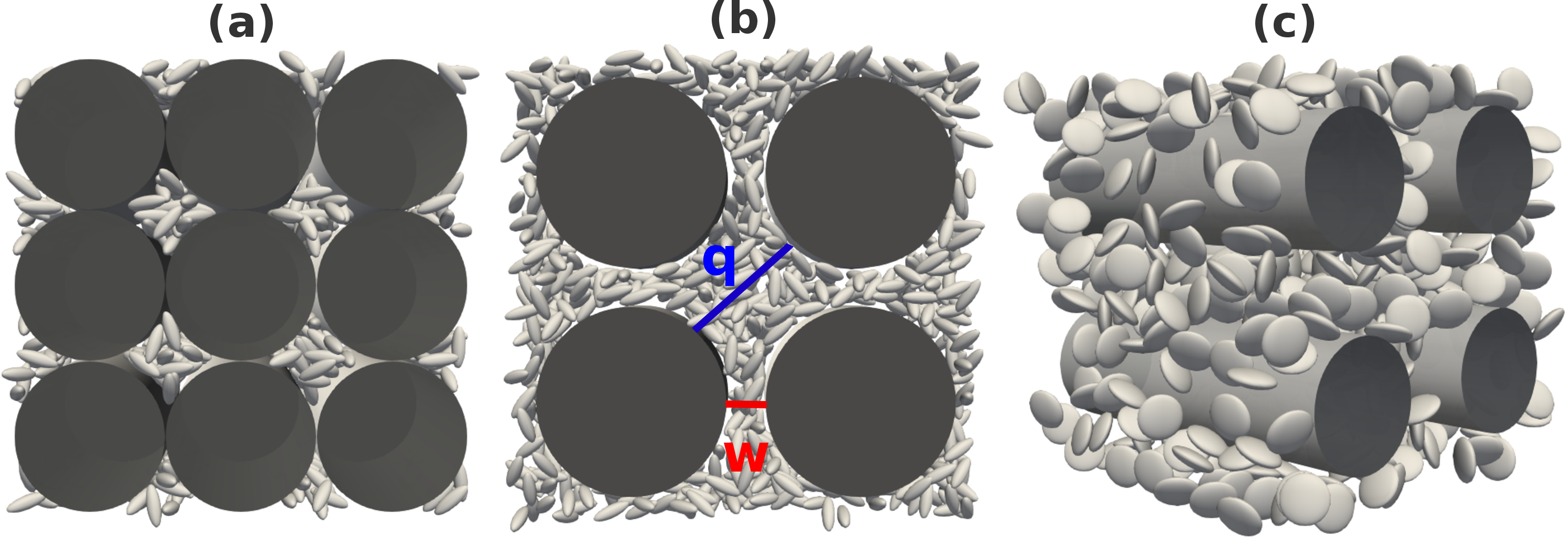}
  \caption{Snap shot of the isotropic system shown at particle concentration $\phi=0.2$ for different obstacle configuration (a) $q=3.0$, $w=0.0$ and $r_\circ=3.6$  (b) $q=7.0$, $w=2.0$ and $r_\circ=5.0$ (c) $q=2.3$, $w=2.0$ and $r_\circ=2$, where $w$ and $q$ are the length measured in the unit of $d_\circ$, as shown in the snap (b). }
\label{fig.4_7}
\end{figure}

\subsection{Dynamics of the system in the presence of obstacles}

The translational diffusion dynamics of a tracer spherical particle in a crowded media, depends only upon the accessible volume \cite{babu_jcp_2008}. Accessible volume is defined as the volume accessible to the center of mass of a tracer particle. That means as we increase the volume fraction, the diffusivity decreases and becomes dependent upon the packing fraction \cite{cristiano_prl_2007, PhysRevE.106.014602}. 
Here we are exploring the dynamics of the ellipsoidal particle far away from the nematic transition. We also did not take into account the hydrodynamic effect. 
When we assume that the fluid particles are very small in size compared to the size of the obstacles, the effect of hydrodynamics on the system remains minimal \cite{goel2009available, kim1992diffusion}. We have also considered the surface of the obstacles to be smooth and featureless \cite{10.1063/1.3651478}. 

The $2$-dimensional diffusivity of the diffusing point particles have been calculated analytically, in the presence of the periodically arranged immobile disks \cite{10.1063/1.4720385}. In order to compare with the analytical results for finite size and shape of the diffusing particles we considered the obstacle configuration in terms of the trap size given as $q$ having a trap opening width of $w$, as shown in the fig \ref{fig.4_7} (b) \cite{PhysRevLett.50.1959}. Here $q$ and $w$ can be expressed in terms of the obstacle radius ($r_o$) and lattice density ($\rho_o$).

\begin{subequations}
\begin{align}
w= \sqrt{\frac{1}{\rho_\circ}} - 2 r_\circ \\
q=\sqrt{\frac{2}{\rho_\circ}} - 2 r_\circ
\label{eq.4_24}
\end{align}
\end{subequations}

The effective translational diffusivity of the point particles in the presence of the $2d$ array of the obstacles is given by, \cite{lifson1962self}

\begin{equation}
D_T^{eff}= \frac{1}{\langle w'(x) \rangle \langle 1/w'(x) D(x)\rangle} 
\label{eq.4_24_a}
\end{equation}

where $w(x)$ is the channel's width, which changes with positions. $w'(x)$ is the derivative of channel width with respect to position\cite{10.1063/1.4720385}. $D(x)$ is the position dependent diffusivity.

When we consider the diffusion coefficient to be a constant (independent of the position) and equal to $D_T^P$, then we can solve Eq. \ref{eq.4_24_a}. The analytical expression for the effective diffusivity due to Fick-Jacob $D_T^{FJ}$ is then given as, \cite{10.1063/1.4720385} 

\begin{equation}
\begin{multlined}
\frac{D_{\text{T}}^{FJ}}{D_T^P} = \frac{1}{\left(1 - \frac{\pi}{4} f_w^2 \right)} \left( \frac{2}{\sqrt{1-f_w^2}} \right. \\
\left. - \arctan\left(\frac{\sqrt{1+f_w}}{1-f_w}\right) - \frac{\pi}{2} + 1 - f_w \right)
\end{multlined}
\label{eq.4_24_b}
\end{equation}

However this assumption is not accurate as in the presence of the obstacles, the volume accessible to the ellipsoidal particles to diffuse will depend on the width of the channel formed between the obstacles. Thus the diffusivity of the ellipsoids $D_T(x)$ will be position dependent. 

Reguera et. al  \cite{PhysRevE.64.061106} considered the short time diffusion of the point particles to be position depended such that $D_T(x)=D_T^P/{[1+w'(x)/4]}^{1/3}$. The effective diffusivity $D_T^{RR}$ is given as,
\begin{equation}
\begin{multlined}
\frac{D_{\text{T}}^{RR}}{D_T^P} = \frac{1}{\left(1 - \frac{\pi}{4} f_w^2 \right)} \\
\frac{1}{\left[ f_w \int_0^{\pi/2} \frac{ (\cos\phi)^{1/3} \, d\phi}{(1 - f_w \cos \phi)} + 1 - f_w \right]}
\end{multlined}
\label{eq.4_24_c}
\end{equation}

Where $f_w$ is given as $(\sqrt{1/\rho_o}-w)/(\sqrt{1/\rho_o}-d_p)$ with $\sqrt{1/\rho_o}$ being the lattice constant for the obstacle configuration and $d_p$, the length of the minor axis. At $f_w = 0$, we have infinitely thin cylinders and the channel opening $w$ between the cylinders reaches its maximum value which is equal to the lattice constant $\sqrt{1/\rho_o}$. At $f_w=1$, the channel width is so small that ($w = d_p$) no ellipsoidal particle can escape, thus the diffusivity perpendicular to the cylindrical axis goes to zero. To compare the diffusivity of a particle with finite volume with that of the point particle model, we have kept $q$ the trap size constant, while changing $f_w$. In other words lattice constant and the radius of the obstacles both are varied, to maintain the desired trap size $q$ and channel width $w$, which are the relevant parameters governing diffusivity of the ellipsoidal particles. 

\begin{figure}
\includegraphics[height=8.8cm,width=8.8cm]{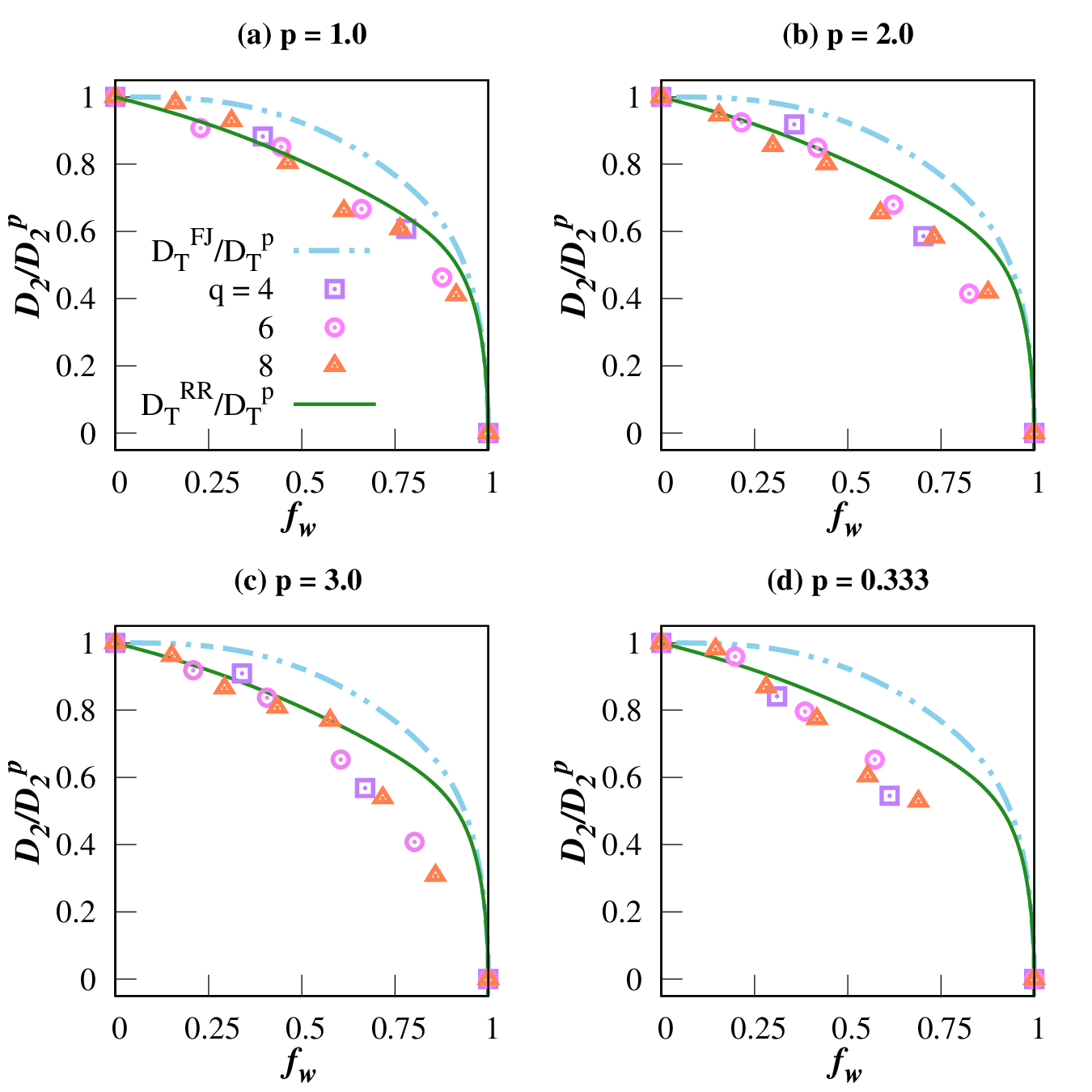}
  \caption{Variation in $2D$ translational diffusivity $D_2/D_2^p$ of single spheroid shown at different cage opening $f_w$, measured in the laboratory frame in the plane perpendicular to the cylindrical axis, for $w=4$ (square), $6$ (circle), $8$ (triangle). (a) For spheroids with $p=1.0$, (b) $2.0$, (c) $3.0$ (rod like particles) and (d) $0.333$ (disk like particles). All the diffusivity have been normalized by $D_2^p$, the diffusivity at $f_w=0$ (for the constant radius of obstacles $\rho_o=0$).
  }
\label{fig.4_8}
\end{figure}

 We calculated the long-time diffusivity of the spheroids in the plane perpendicular to the axis of the cylinder by changing $f_w$ at different constant $q$ values as shown in Fig \ref{fig.4_8}. The resulting $2$-dimensional translational diffusivity $D_T^2$ for the single particle at different $f_w$ is normalized by the $D_2^p$ which is also $2d$ diffusivity at $f_w=0$ for a particular trap size $q$. $D_T^2$ for spheres (Fig. \ref{fig.4_8} (a)) agrees with the $D_T^{RR}$ (solid green line), however $D_T^{FJ}$ over estimates the diffusivity of the ellipsoidal particles. When we increase the anisotropy of the ellipsoidal particle by changing $p$, we observe significant deviation from the analytical calculation see Fig. \ref{fig.4_8}(b), Fig. \ref{fig.4_8}(c) and Fig. \ref{fig.4_8}(d). For the prolate particles the deviation from the analytical calculation happens when $f_w>0.5$ as can be observed in Fig. \ref{fig.4_8}(b) and Fig. \ref{fig.4_8}(c). It can also be observed that when the trap size $q$ is smaller the deviation from analytical calculation happens at lower $f_w$. For the disk like particles Fig. \ref{fig.4_8}(d) the deviation from the analytic calculation happens at a smaller $f_w$ than for the needle like particles. Also note that in the absence of the obstacles both the system $p=0.333$ and $3.0$ shows almost the same diffusivity \cite{PhysRevE.106.014602}. 
 
\begin{figure}
\includegraphics[height=8.8cm,width=8.8cm]{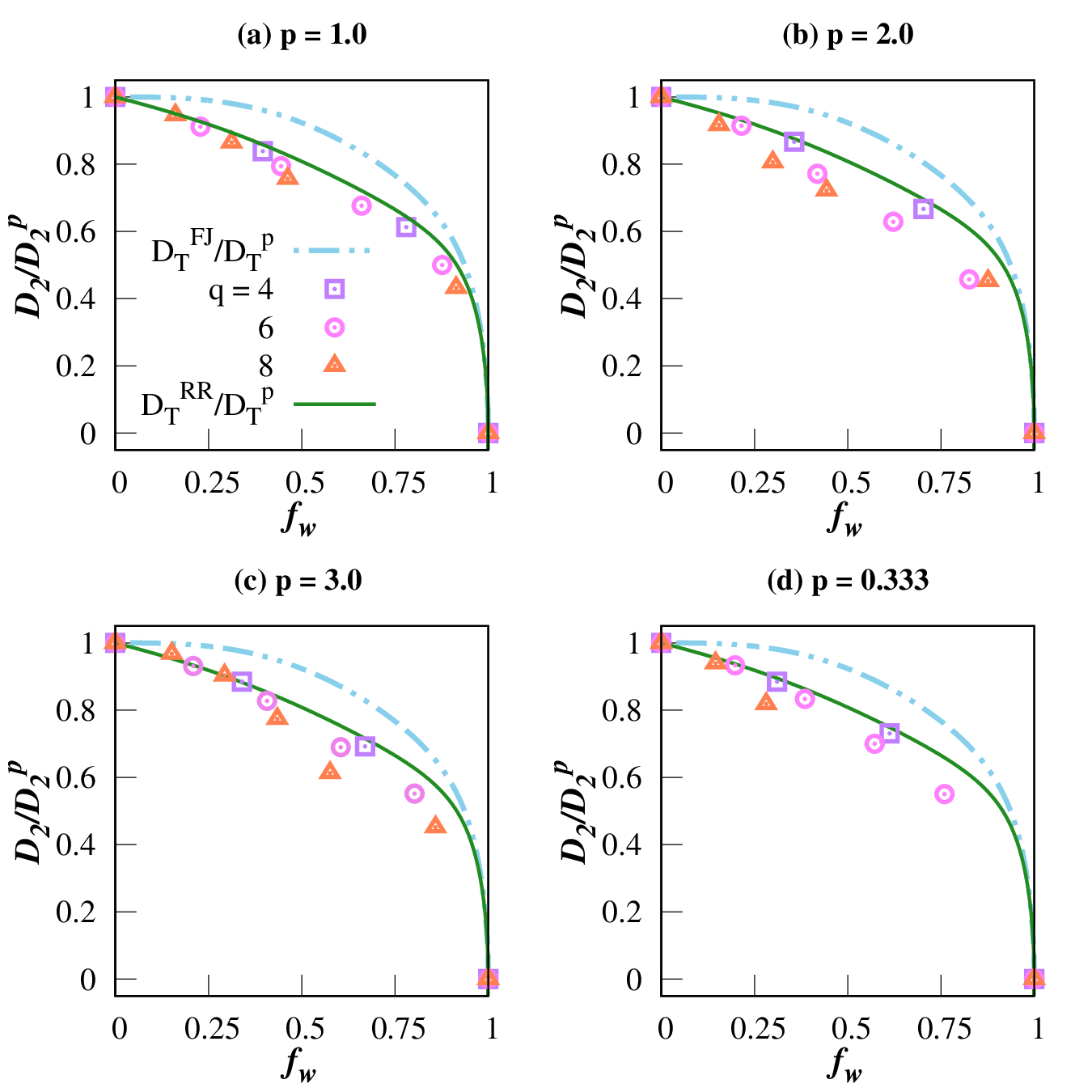}
  \caption{Variation in $2D$ translational diffusivity $D_2/D_2^p$ of spheroids at finite concentration ($\phi=0.2$), shown at different cage opening $f_w$, for $w=4$ (square), $6$ (circle), $8$ (triangle). (a) For spheroids with $p=1.0$, (b) $2.0$, (c) $3.0$ and (d) $0.333$ }
\label{fig.4_9}
\end{figure}

 
 There are no effect of the volume fraction of the particles have been considered in the effective diffusivity of the particles in the presence of the obstacles. To compare with the expression of the diffusivity given for the point particles, we calculated $D_2$ and normalized with the $D_2^P$, which is the maximum $D_2$ possible at a constant $q$ (at $f_w$=0) for a given finite volume fraction $\phi=0.2$. For the spherical particles, the diffusivity follows the analytical curve as in the case of single particle diffusivity as shown in the Fig. \ref{fig.4_9} (a). However, for the ellipsoidal particles also, we have less deviation from the analytical curve in compare to the single particle diffusivity, as shown in the Fig. \ref{fig.4_9}(b), \ref{fig.4_9}(c) and \ref{fig.4_9}(d) calculated  for $p=2.0$, $3.0$ and $0.333$ respectively. It indicates that for the particles at finite volume fraction in the isotropic phase the caging effect due to the obstacles remains the same regardless of the shape of the particle.

\begin{figure}
\includegraphics[height=8.8cm,width=8.8cm]{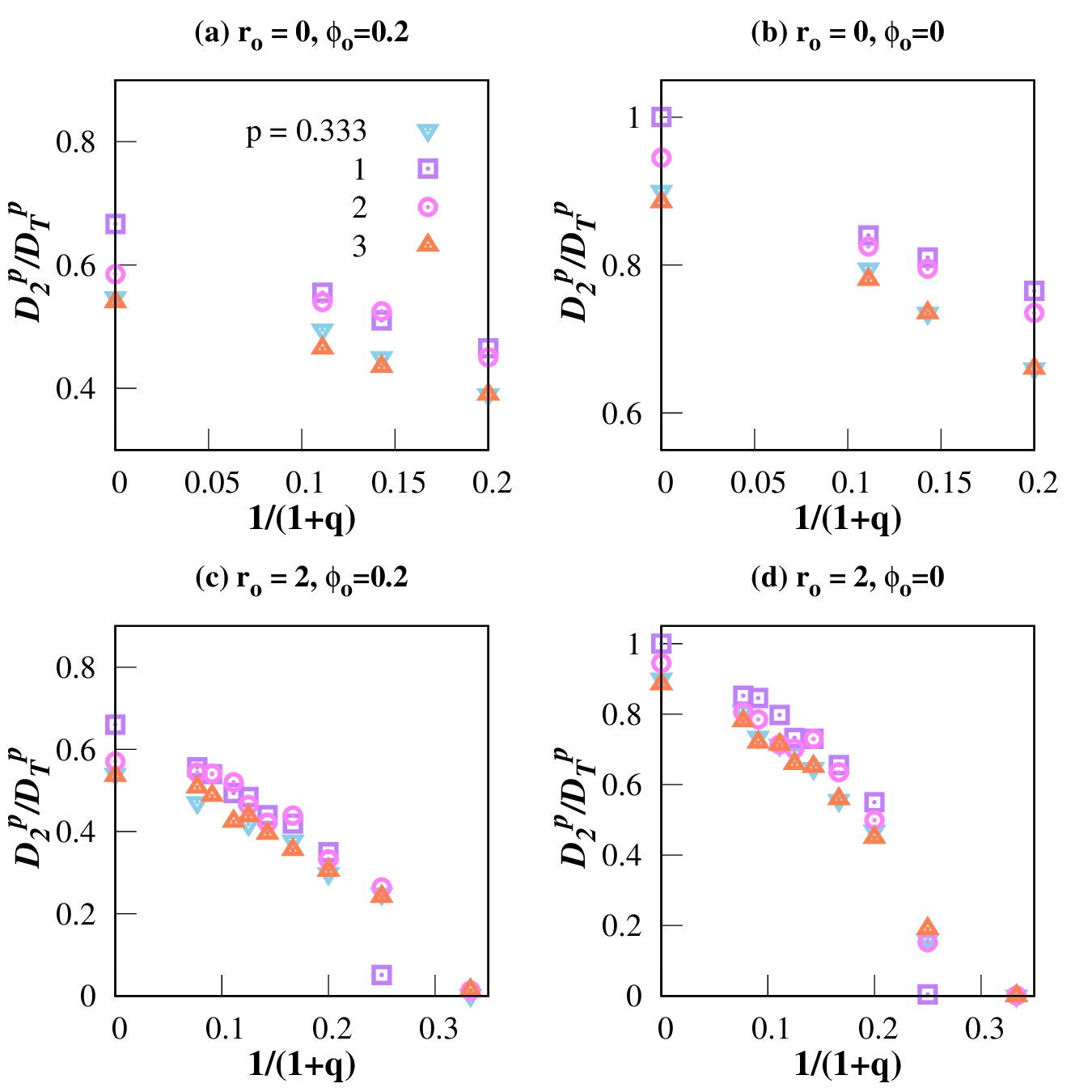}
  \caption{Variation in the $D_2^p/D_T^p$ with $1/(1+q)$, where $D_2^p$ is the diffusivity shown for the cylinders with a constant and infinitely thin radius ($\rho_o=0$). (a) At finite particle's concentration and (b) at infinitely dilute concentration. Variation in $D_2^p$  also shown for the finite radius ($\rho_o=0$)(c) at finite particle's concentration and (d) at infinitely dilute concentration. Where all the diffusivity values have been scaled by $D_T^p$, the diffusivity of single spheroids in the bulk phase.}
\label{fig.4_10}
\end{figure}

For a particular constant $q$ value, the diffusivity will be maximum $D_2^p$ at $f_w=0.0$ (at constant $r_o=0$). Fig. \ref{fig.4_10} shows variation in $D_2^p$ with $q$, where the x-axis is scaled as $1/1+q$, which varies between $0$ to $1$, as $q$ varies between $\infty$ to $0$. $q=\infty$ represents a system without obstacles. In the larger trap size regions the diffusivity varies almost linearly regardless of their shapes(Fig. \ref{eq.4_10} (a) and (b)). One difference between the Lorentz gas model and the finite size particle is that $D_2^p$ of the Lorentz gas model would be maximum and same at all the $q$ values \cite{PhysRevLett.50.1959} for $f_w=0$. To find out the variation in the $D_2^p$ in the small trap size regions (comparable with particle's size), we calculated diffusivity at $r_o=2.0$ (Fig. \ref{eq.4_10} (c) shows $D_2^P$ at finite concentration and \ref{eq.4_10} (d) for single particles ). For the spherical particles (square points) diffusivity drops abruptly. While the drop in the diffusivity of the anisotropic particles remains linear almost throughout the whole $q$ regions.

\begin{figure}
\includegraphics[height=6.3cm,width=9cm]{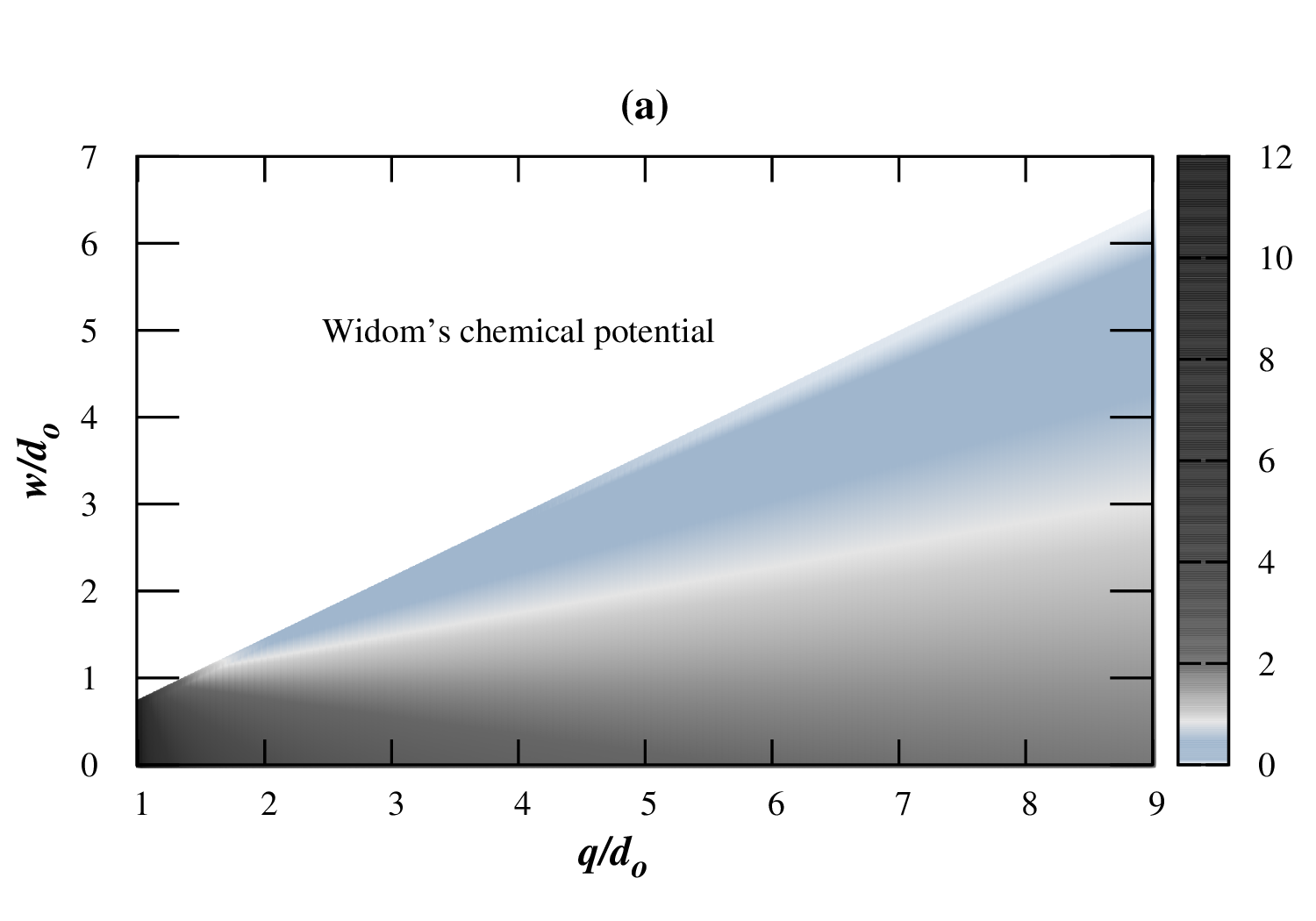}
  \caption{Widom's chemical potential is shown by the colored map in $q-w$ plain. }
\label{fig.4_11}
\end{figure}

\begin{figure}
\includegraphics[height=13.2cm,width=8.8cm]{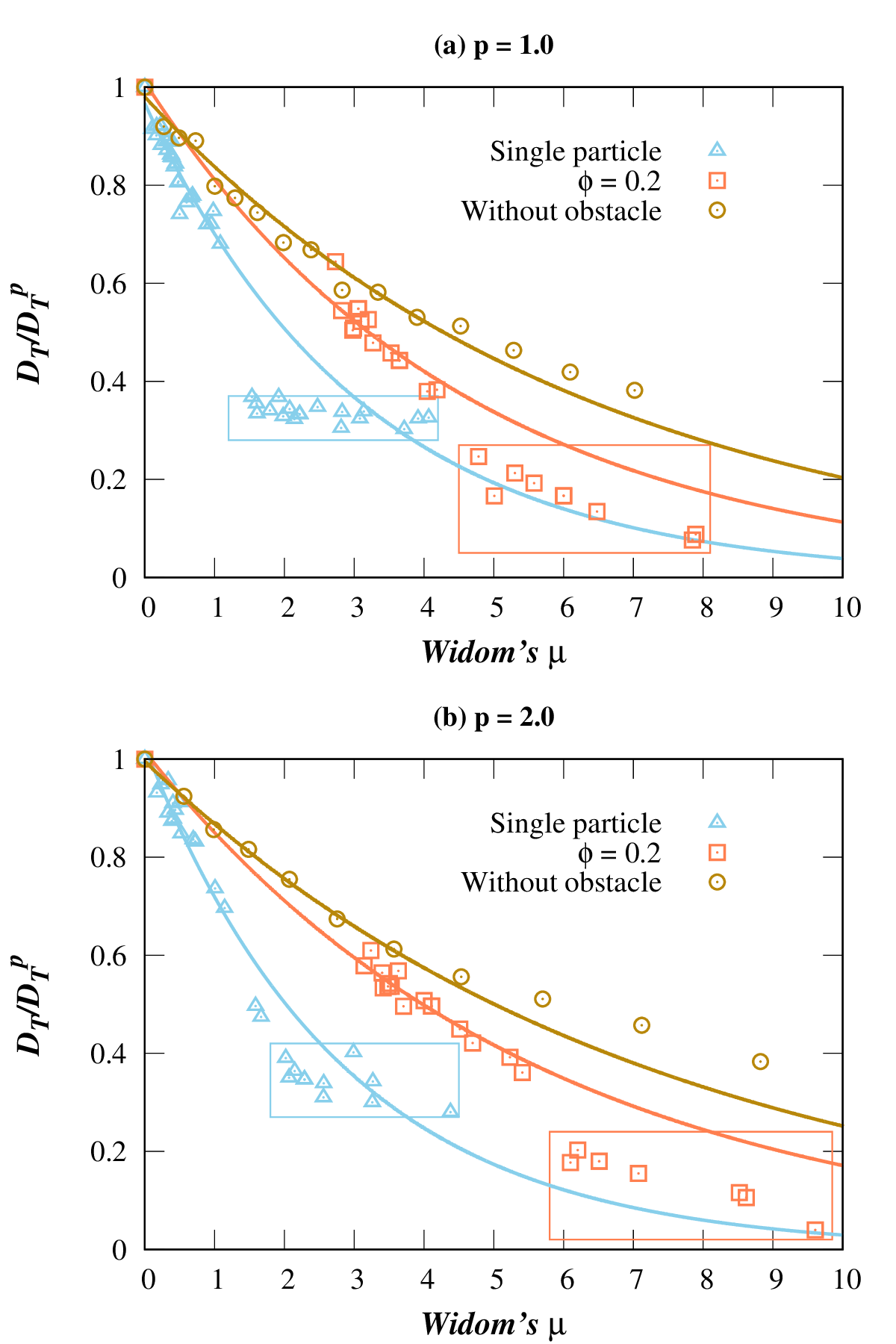}
  \caption{Translational diffusivity $D_T/D_T^p$ normalized by the diffusivity of the particle in infinite dilution $ D_T^p$ is shown as the function of the Widom's chemical potential for (a) spherical particles and (b) prolate particles. Blue triangles represent the single particle diffusivity within the presence of the obstacles. Coral squares represent the diffusivity at finite particle concentration with $\phi=0.2$, compared with the diffusivity at different particle concentrations $\phi$ in the absence of the obstacles represented by dark-golden circles. Solid lines are the exponential fits. Points inside the square bracket represent the system with $2d$ confinement ($s<1.0$), where the particles can only move in the $1d$ pour confined by the cylindrical obstacles.}
\label{fig.4_12}
\end{figure}

\begin{figure}
\includegraphics[height=13.2cm,width=8.8cm]{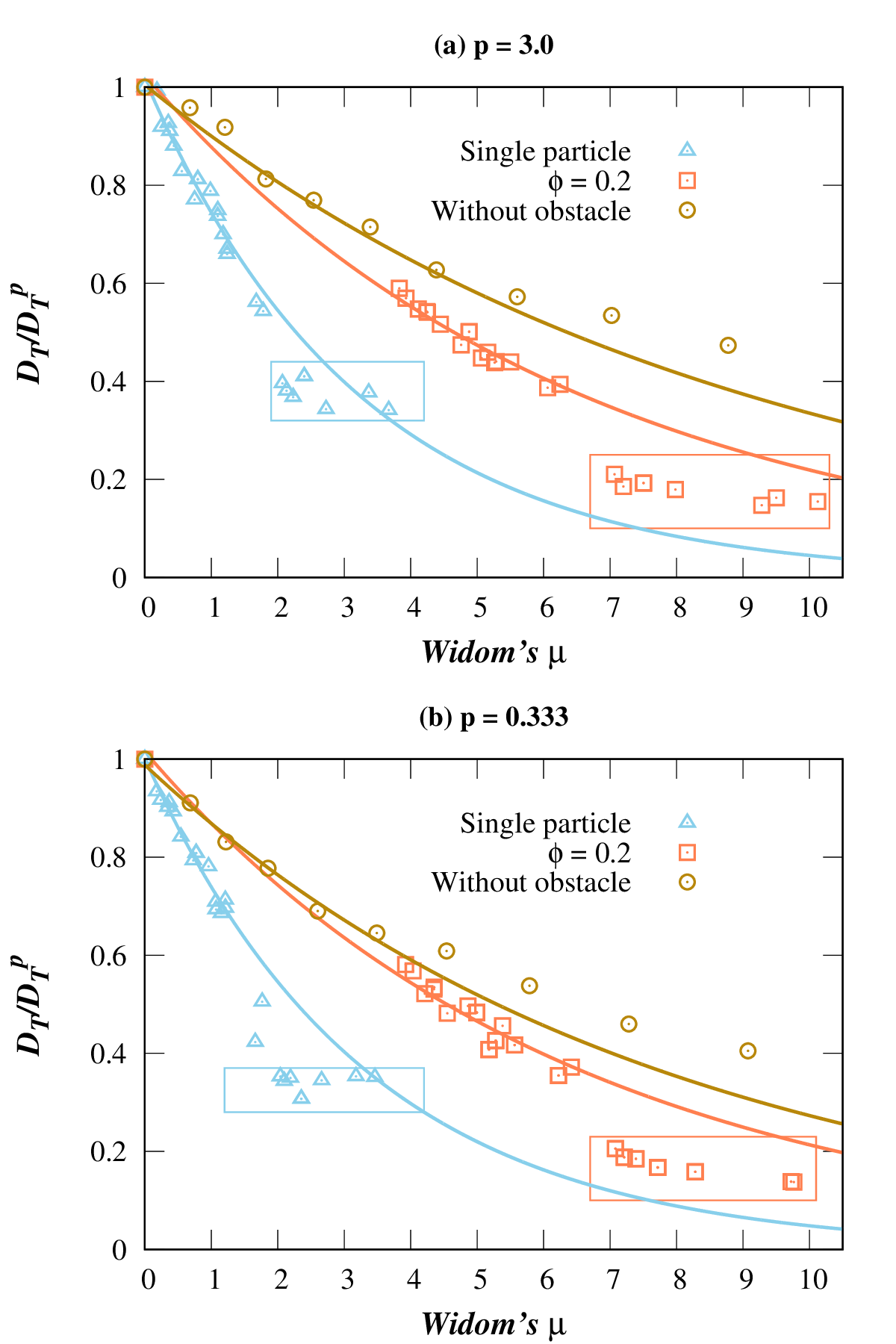}
  \caption{Translational diffusivity $D_T/D_T^p$ normalized by the diffusivity of the particles in infinite dilution $ D_T^p$ is shown as the function of the Widom's chemical potential for highly anisotropic particles (a) prolate ($p=3.0$) and (b) oblate ($p=0.333$). Blue triangles represent the single particle diffusivity within the presence of the obstacles. Coral squares represent the diffusivity at finite particle concentration with $\phi=0.2$, compared with the diffusivity at different particle concentrations $\phi$ in the absence of the obstacles represented by dark-golden circles. Solid lines are the exponential fits. Points inside the square bracket represent the system with $2d$ confinement ($s<1.0$), where the particles can only move in the $1d$ pour confined by the cylindrical obstacles.}
\label{fig.4_13}
\end{figure}

The dynamics of the ellipsoidal particle is related to the volume accessible to the center of mass of the ellipsoidal particles also known as accessible volume, due to the finite size of the obstacles and the ellipsoids. If the accessible volume is small it will lead to caging, and dimensional confinement. The diffusivity of the ellipsoidal particles goes to zero when the channel width is close to the ellipsoidal particle's size. The accessible volume is related to the chemical potential of the hard sphere particle system \cite{babu_jpcb_2008} given as $\mu/k_BT=-\ln{\frac{\phi_a}{\phi}}$, where $\mu$ is the chemical potential of the hard sphere particles at volume fraction $\phi$ and having accessible volume $\phi_a$. In Fig. \ref{fig.4_11} we have plotted the chemical potential calculated using Widom's particle insertion method for different $\phi$ with varying $w/d_\circ$ and $q/d_\circ$ at different $\phi_o$. For constant $w/d_\circ$ values, $\mu/k_BT$ decreases with an increase in $q/d_{\circ}$ and becomes constant in the large $q/d_\circ$ regime corresponding to large accessible volume. When $w<d_\circ$ and $q<d_{\circ}$, $\mu/k_BT$ increases it becomes increasingly difficult to access free volume for the ellipsoidal particle. Higher $\mu/k_BT$ represents lower accessible volume, which leads to the decrease in diffusivity of the particles.

Experimentally diffusivity of the tracer particles has been shown to fall exponentially with the increase in the concentration of heavy globular proteins in the aqueous solution \cite{banks2005anomalous}, due to the decrease in accessible volume for the tracer particle. To show the effect of the confinement, excluded volume, $\phi$ and $\phi_a$ on the translational diffusivity, we have shown $D_T/D_T^p$ as the function of Widom's chemical potential $\mu/k_BT$ for the spherical ($p=1.0$ Fig \ref{fig.4_12} (a)) and spheroidal particles (prolate, $p=2.0$ Fig \ref{fig.4_12} (b)). At this $p$ we will always have isotropic fluid phase irrespective of the $\phi$. The  diffusivity of the system with respect to $\mu/k_BT$ is shown at finite spheroid concentration and infinite spheroid dilution (single particle) in comparison with the diffusivity of hard spheroid at different $\phi$ without the presence of obstacles. We can observe that a sudden sharp fall in diffusivity of the spherical particle and prolate particle as $w/d_\circ$ falls below $1.0$ (size of the particle) as shown in Fig. \ref{fig.4_12} (a) and Fig. \ref{fig.4_12}(b) respectively. Interestingly in the case of single particle the diffusion coefficient becomes independent of the obstacle concentration for both spherical and prolate particle. The reason being the sphere and the prolate particle can only diffuse along one dimension which is parallel to the axis of the cylinder. The diffusion coefficient $D_T/D_T^p$ for a spherical and the prolate particle along one dimension in simulations is $\sim 0.35$, which is the approximate value we observe for spheres.  
At finite concentration ($\phi=0.2$), the diffusivity keeps decreasing with increasing $\mu/k_BT$ due to the increase in $\phi$, as the effective accessible volume fraction decreases due to the excluded volume between the two ellipsoidal particles. Also with the increase in the radius of the cylindrical obstacle,  all the ellipsoidal particles may get aligned in a column \cite{zarif2020mapping}. This is the minimum energy configuration for a single prolate particle in the presence of the cylindrical obstacle. Thus for a finite volume fraction once the ellipsoids are aligned then the diffusion is only possible along one dimension which is along the direction of the cylindrical axis. We observe that diffusivity of the ellipsoidal particle at $\phi=0.2$ (Fig \ref{fig.4_12} (a) and \ref{fig.4_12} (b))  follows the fitted exponential curve, which fall suddenly to lower value. This is again due to the one dimensional confinement of the ellipsoidal particle. Compared to the single particle diffusion which was a constant after confinement along one dimension for the $\phi=0.2$ even after confinement the diffusion coefficient falls due to the finite volume fraction of the ellipsoids. In all the cases, the diffusivity in the presence of the obstacles remains lower than the diffusivity in the spherical particle at finite $\phi$ without the obstacle. 


\begin{figure}
\includegraphics[height=8.8cm,width=8.8cm]{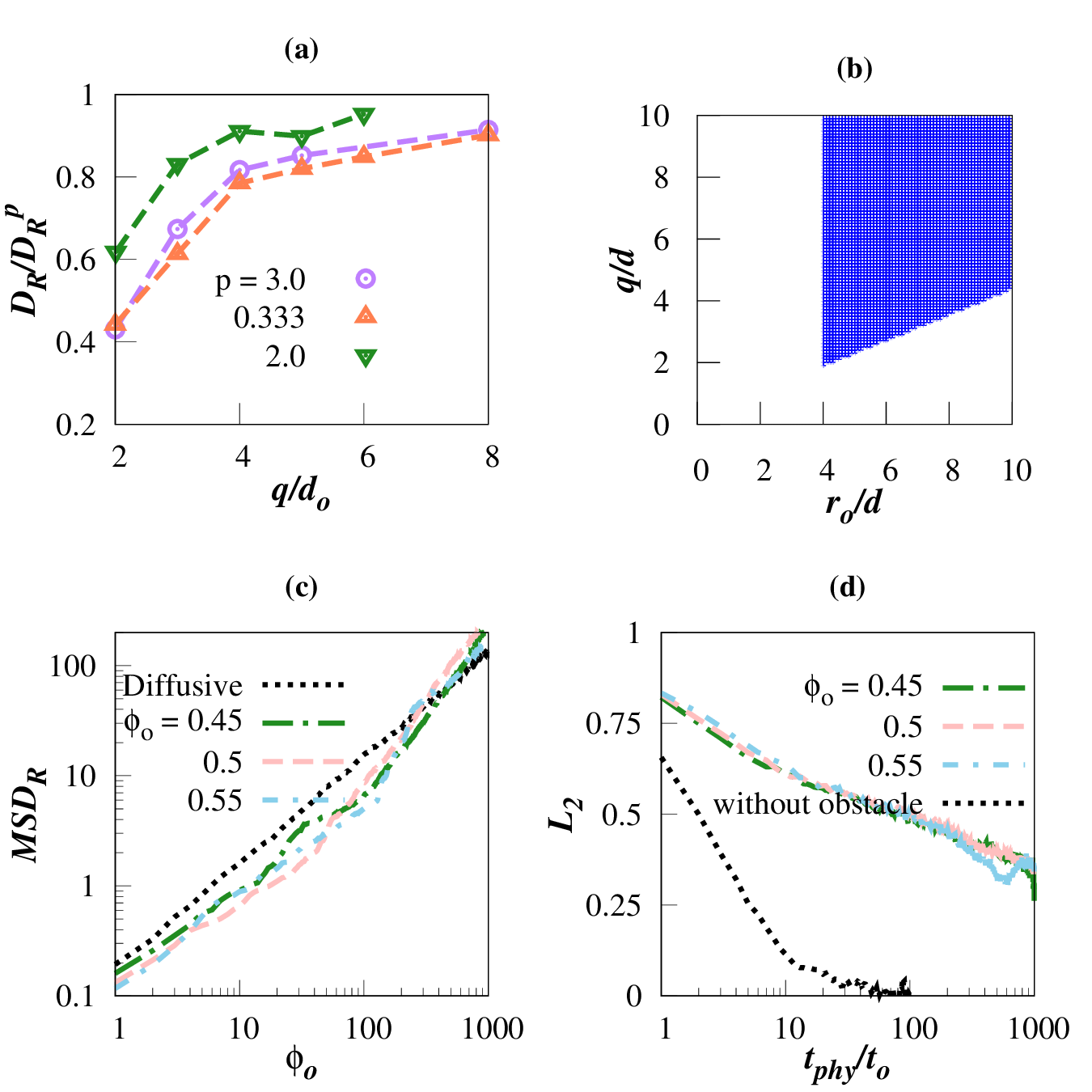}
  \caption{(a) Rotational diffusivity $D_R/D_R^p$ is shown with the change in obstacle configuration, for $p=3.0$ (circles), $p=0.333$ (upper triangle) and $p=2.0$ (lower triangle), at particle concentration $\phi=0.2$. (b) Blue region shown in the $q-r_o$ plane represents the configuration of obstacles, where the disk like particles shows diffusive behaviour at long time scale despite being in nematic phase (for $\phi_o>0.425$). (c) $MSD_R$ is shown for the oblate particles ($p=0.25$) in nematic phase ($\phi=0.45$), at different obstacle concentrations $\phi_\circ$,  for a large obstacle size $r_\circ=5.0$. The black line is a guide to the perfectly diffusive behavior. (d) Decay of $L_2$ with time is shown. Long dashed lines represent the different $\phi_\circ$ as given in the legend, compared with the decay of $L_2$ for $D_R/D_R^p=0.2$ in the absence of obstacles calculated at $\phi=0.45$ (short dashed black line)}
\label{fig.4_14}
\end{figure}

To show the effect on the dynamics of highly anisotropic particles, we took oblate and prolate particles with $p=3.0$ and $p=0.333$, respectively. It should be noted that, the presence of the cylindrical obstacles do not change the isotropic-nematic boundary in the $\phi-p$ plain. Fig \ref{fig.4_13} (a) and Fig \ref{fig.4_13} (b) show the variation in $D_T/D_T^p$ of both prolate and oblate particles, respectively, as a function of $\mu$, calculated at $\phi=0.2$. At this $\phi$ the systems corresponding to both prolate and oblate remains isotropic for all the points shown in the Fig \ref{fig.4_13} (a) and Fig \ref{fig.4_13} (b). The translational diffusivity behaves in the same way as a spherical particle system. The ellipsoidal system still shows a sudden drop with $\mu$ as the diffusivity perpendicular to the cylinder axis drops to zero as discussed earlier for the spherical system.

Unlike the translational diffusivity which depends dominantly on the channel width $w/d_\circ$, the rotational diffusivity is found to depend dominantly on the trap size $q/d_\circ$. At large $q/d_\circ$ the rotational diffusivity remains decoupled with the translational diffusivity in the isotropic phase. Fig \ref{fig.4_14} (a) shows the variation in rotational diffusivity $D_R/D_R^p$ with $q/d_\circ$. $D_R/D_R^p$ is shown to stagnate quickly with the increase in $q/d_\circ$ reaching the diffusion coefficient of the bulk system in the absence of obstacles ($D_R^p\sim 0.33, 0.56, 0.66$, for $p=3.0, 0.33, 2.0$ respectively calculated  at $\phi=0.2$) for both prolate as well as oblate system. However, in the nematic phase the disk like particles show strong coupling between translational and rotational diffusivity. In the absence of obstacles, the rotational motion in the nematic phase shows sub-diffusive behavior in the long time limit.  However, in the presence of the obstacles with with high $r_\circ/d>4$, $D_R/D_R^p$ of the oblate particles present in the nematic phase near the obstacle's come out of the dynamic arrest configuration. As it has been shown in the $g(\theta)$ calculation (Fig. \ref{fig.4_6}(a)), with higher radius the obstacles structure get frustrated locally, providing space for the flipping of the particle. The region over which we observe the anomalous behavior is observed is shown as the shaded region in the Fig \ref{fig.4_12} (b) in the $q/d-r_\circ/d$ plane, where the disk like particles with $p=0.25$ were considered. 

The oblate particle close to the obstacles flips along the major axis, contributing to the higher value of diffusivity. We observe the dynamics deviating away from being sub-diffusive and enters the super diffusive phase. 
Rotational mean-square-displacement ($MSD_R$) \cite{doi:10.1073/pnas.1203328109} is shown in the Fig \ref{fig.4_12}(c), where we observe a shift from the short time sub-diffusive to long time super-diffusive regime. As for the $MSD_R$ only the rotating particle contributes because all the other particle is in the nematic phase and hence show sub-diffusive behavior. To show that this is a local effect we have calculated the orientational self-correlation $L_2$. given as, 

\begin{equation}
L_2(t)=\frac{1}{N}\sum_{i} \frac{1}{2} (3\hat{n}_i(\tau+t).\hat{n}_i(\tau)-1)
\end{equation}

where $n_i(t)$ is the orientation of the $i^{th}$, particle at a time $t$. We have shown the decay of the $L_2$ black dashed line in Fig \ref{fig.4_14}(d) for a system without obstacle, with nearly the same $D_R/D_R^p=0.19$ (at $\phi=0.4$). $L_2$ decays to $0$ at a $t_{phy}>20.0$. In the presence of the obstacles where the bulk is found to be in the nematic phase and shows $D_R/D_R^p>0.19$,  $L_2$ can be observed not decaying (colored dashed lines in Fig \ref{fig.4_14}(d)). Thus globally the system is in nematic phase which leads to a decline in the rotational dynamics. However, in the presence of obstacles, the particles are flipping locally and contributing to the super-diffusive behavior. In this way system maintains the nematicity despite few of the particles being out of the arrested rotational dynamics phase.

\section{Conclusion}
Experimentally, numerous works have been done by confining the particles in various geometries. In almost all these works, it has been shown that the shape anisotropy of the particles enhances the motion of the self-propelled particle. However, in the passive particle system, dynamics is governed by the structure of the fluid. The theoretical prediction of diffusivity for the point particles is modified to incorporate finite volume particle which agrees with our simulation results. Although with the increase in shape anisotropy, the diffusivity deviates from the proposed theoretical prediction. At finite volume fraction, the dynamics of the system become dependent upon structural properties. The structural property of the spheroidal particle as result of the introduction of obstacles in the system influences the dynamics. As shown in the case of oblate particles, the rotational diffusivity couples with translational diffusivity resulting in the flipping motion, leading to super diffusivity behaviour. In the present study for a low particle concentration regime, the translational diffusivity of the anisotropic particles behaves nearly the same way as the system of spherical particles. For a single particle as well as for finite $\phi$, diffusing in the cylindrical pores with extreme confinement, the long-time translational diffusivity behaves like a one dimensional system, where diffusivity scales to one by a third of the diffusivity of the bulk system. 

In the presence of obstacles, the nematic phase has been observed to show exciting results. We have shown the nematic director being controlled by the presence of cylindrical obstacles. Kinetics of the evolution of the particles' alignment showed that the spheroidal particles preferred to align quickly along the axis of the cylindrical obstacles. We hope the present study would help to design system where the nematic alignment can be controlled by surface effects of the obstacles.  

\section*{Acknowledgements}
VAV would like to acknowledge UGC-CSIR and institute for funding this research. 

\bibliographystyle{apsrev4-1}
\bibliography{Obstacles}

\end{document}